\documentstyle[12pt]{article}
\advance\oddsidemargin	by -1.5in
\advance\evensidemargin by -1in
\newcommand\beq{\begin{equation}}
\newcommand\eeq{\end{equation}}
\newcommand\beqn{\begin{eqnarray}}
\newcommand\eeqn{\end{eqnarray}}

\title{Decisive search for a diquark-antidiquark meson with
hidden strangeness}
\author{Boris Z. Kopeliovich$^{1}$ and	Enrico Predazzi$^{2,3}$\\
\\
$^1$ \small\it Joint Institute for Nuclear Research\\
\small\it Head Post Office, P.O. Box 79, 101000 Moscow,
Russia\\
\\
$^2$ \small\it Physics Department, Indiana University\\
\small\it Bloomington, IN 47401 - U.S.A.\\
\\
$^3$ \small\it On leave from Dipartimento di Fisica Teorica,
Universit\`a di Torino\\ \small\it and INFN, Sezione di Torino,
I-10125, Torino, Italy}

\begin{document}
\maketitle
\begin{abstract}
Diquark-antidiquark states are expected to exist as a natural
complement of mesons and baryons. Although they were
predicted long ago, and some candidates were found
experimentally, none has, as yet, been reliably identified. We suggest
that the search for the so-called $C(1480)$-meson in reactions such as
photoproduction $\gamma N\rightarrow\phi\pi N$ and $K N \rightarrow \phi
\pi \Lambda$ should provide a
decisive  way to settle this issue. Estimates of the cross sections
are given using
present experimental information on the C-meson and
assuming its diquark-antidiquark structure.
Sizable cross sections are predicted (of the order of 0.1 $\mu$b for
photoproduction and of the order of 0.1 mb for $KN$ at the maximum
with an insignificant background).
Failure to find this kind of signal would imply that the C-meson is {\it not}
a diquark-antidiquark state.
\end{abstract}
\newpage

{\it 1. Historical perspective}: within the quark model, ordinary mesons
are $q-\bar q$ states {\it i.e.} color singlets made of (spin 1/2) elementary
color $\{3\}-\{\bar 3\}$ called quarks. Very soon after the quark model
was proposed, it was also suggested [1] that four quark or exotic states
made of diquark-antidiquark pairs should exist. This is in keeping with
the notion that the color singlet $qqq$ baryons states can also, to some
extent and in some circumstances, be viewed as objects composed of a
quark-diquark systems. In both cases (exotic mesons and baryons), what
makes the parallelism with ordinary mesons stringent, is that
the diquarks act again as (spin 0 or 1) color antitriplets (recall that a
two-quark or diquark system can be decomposed as an antitriplet plus a
sextet or $3 \otimes 3 = \bar 3 \oplus 6$).

While most authors would consider diquarks as useful
and simple ways of describing complicated non-perturbative phenomena in
some intermediate energy domain, they do indeed show some elementarity.
For an updated review on diquarks, see Anselmino et al. [2], where an
extended literature to earlier references can also be found.

{\it 2. Decay channels}: $q-\bar q$ mesons decay mostly into other mesons
by creating new $q-\bar q$ pairs from the vacuum. The same mechanism
explains also the decay of baryons into baryons plus mesons.
Such a mechanism can also provide the decay of a
$(\bar q\bar q)-(qq)$ state to baryon-antibaryon (plus mesons) if its
mass is above the production threshold. However many states are
predicted below the threshold, and may therefore only decay into mesons.
The new specific mechanism which must be at work in this case
is rearrangement of the quarks from $qq$ and $\bar q \bar q$ into two
$q \bar q$ states. In this case, however, as in so many other hunts for
exotics, the essential problem is to find a signature which
provides a reliable identification .

It was suggested long ago [3] that a promising signature could be
associated to a
diquark-antidiquark state with hidden strangeness,
$(\bar q\bar s)-(qs)$. As a result of quark rearrangement, such a
state could only decay, presumably with about equal probabilities,
either into $K\bar K$ or to $\phi\pi$ (additional
pions are of course possible). In particular, the latter channel is
expected to be suppressed by at least two orders of magnitude by the
OZI rule, if the parent meson has a $q-\bar q$ or $s-\bar s$ structure
instead of the diquark-antidiquark one. So, the observation of a
state, decaying into $\phi\pi$ with {\it non-suppressed branching
ratio for this channel}, would indeed be a reliable signature of its
$(\bar q\bar s)-(qs)$ structure.

{\it 3. Experimental situation}: A candidate for such a state (named
C(1480) in the literature) was found in Serpukhov [4] in the reaction
$\pi^-p\rightarrow\phi\pi n$ with a 32.5 GeV/c pion beam and with a
cross section
\beq
\sigma(\pi^- p \rightarrow Cn) BR(C \rightarrow \phi \pi^0) = 40 \pm 15 nb.
\label{1}
\eeq

Unfortunately, in this experiment, only the total width,
$\Gamma_{tot}=130 \pm 60\;MeV$ was established. The presence of the
unknown vertex $C\rightarrow\pi\pi$ in this process prevents any
possibility to fix the branching ratio of the $\phi\pi$ channel.
As a consequence, the interpretation of this
state is still questionable.

In addition, this state $C(1480)$ was searched for by the ASTERIX
collaboration in antiproton annihilation at rest, $\bar
pp\rightarrow\phi\pi\pi$. No signal was found, and only an upper
limit was put on the cross section of C-meson production [5].
However, in a situation when the production cross section
cannot be reliably predicted (which is also the case for the Serpukhov
experiment) there could be no contradiction between the two
experiments.

The only way to settle the issue is to search for the state under
consideration in processes with predictable cross section; this
would also provide the possibility to measure the partial width for
the decay into $\phi\pi$.

{\it 4. Photoproduction of $C(1480)$}: In this paper we suggest
to search for the state $C(1480)$ in the
reactions $\gamma N\rightarrow\phi\pi N$. The
diagram, corresponding to photoproduction is shown
in fig.1. We use here the
vector dominance model, and assume pion exchange. According to
previous consideration, the partial width $C\rightarrow\phi\pi$
should be roughly a half of the total width, if the C meson is a
$(\bar q\bar s)-(qs)$ state. In such a case all vertices of this
diagram are known, and the cross section can be reliably predicted
as,
\beq
\frac{d\sigma}{dt\;dM^2}=\frac{\alpha}{f_{\phi}^2}\;
\frac{g_{\pi NN}^2}{4\pi}\;
\frac{-t}{(m_{\pi}^2-t)^2}\;
\frac{M^2}{s^2}\;
\sigma^{\pi\phi}_{el}(M)\;e^{R^2t}
\label{2}
\eeq
The $\pi\phi$ elastic scattering cross section at the c.m.
energy M, $\sigma^{\pi\phi}_{el}(M)$ has two components,
\beq
\sigma^{\pi\phi}_{el}(M)=
\sigma^{\pi\phi}_{P}(M)+
\sigma^{\pi\phi}_{C}(M)
\label{3}
\eeq
where $\sigma^{\pi\phi}_{P}$ is a non-resonant background provided by
Pomeron exchange (other Reggeon exchanges are strongly suppressed by the OZI
rule and will be ignored). This contribution can easily be estimated
using the additive quark model with a reliability which is largely sufficient
for our present purposes. For the total cross section we find
$\sigma^{\pi\phi}_{tot}\approx
8$ mb.
The same estimate follows from VDM analyses of data on
$\phi$ photoproduction on nuclei [6]. Then, we use
$\sigma^{\pi\phi}_{P}=(\sigma^{\pi\phi}_{tot})^2/16\pi
B^{\pi\phi}_{el}$ to estimate the elastic cross section. Using
$B^{\pi\phi}_{el}\approx 6$ GeV$^{-2}$,
we get as an energy-independent
background $\sigma^{\pi\phi}_{P}(M)\approx 0.6$ mb. \smallskip

For the resonant contribution coming from the C(1480) meson
to $\sigma^{\pi\phi}_{el}(M)$ we use the following form
\beq
\sigma^{\pi\phi}_{C}(M)=4\pi\;\frac{k_f}{k_ik_0^2}\;
\left[1+4\;\frac{(M-M_C)^2/}{\Gamma_{tot}^2}\right]
\label{4}
\eeq
where we have taken $\Gamma_{\phi\pi}\approx1/2\Gamma_{tot}$.
$k_i, k_f$ are the $\phi - \pi$ c.m. momenta in initial and
final states respectively. The former is calculated for massless
$\phi$ and $\pi$. For $k_0$ we take $k_0=k_f(M=M_C)$. This combination of
momenta takes into account the threshold behavior of the partial width
$\Gamma_{\phi\pi}$ and corrections to the flux factor for the
off-mass-shellness of $\phi$ and $\pi$ in the initial state.
As for the other parameters appearing in (4),
we use standard values: $f_{\phi}^2/4\pi\approx
14$; $g_{\pi NN}^2/4\pi\approx 15$; $R^2\approx 3.3$ GeV$^{-2}$
[6,7].

We integrate (2) from $t_{min}$ to $t_{max}$, which are the values of the
minimal and of the maximal 4-momentum transfer squared in $\phi - \pi$
scattering, taking into account that the particles are off-shell
in the initial state. This provides a correct energy
behavior of the cross section at threshold.

The result for the energy dependence of the cross section of C-meson
photoproduction is shown in fig.2 versus photon energy. The cross
section shows a maximum at $E_{\gamma}\approx 4.5$ GeV
which results
from the interplay of the near threshold growth and of the
decrease with energy as $s^{-2}$ of the pion-exchange contribution.

The $\phi\pi$ mass distribution is shown in fig.3 at $E_{\gamma}
=4$ GeV. In spite of the phase factor $M^2$ in (2) the
$M^2$-distribution looks pretty symmetrical as a results of the
compensating threshold behavior at large $M$. The
background of the Pomeron-exchange elastic cross section,
$\sigma^{\phi\pi}_P$, is also included but its effect, however, proves
to be negligibly small.

 \smallskip
Our result deserves some comment, given that it
corresponds to the somewhat surprisingly large cross section
of about $100$ nb whereas one would naively have expected a strong
suppression due to the OZI rule had the C-meson been just an ordinary
meson. It is just the alleged diquark-antidiquark
structure of the C-meson which removes the
suppression. Note that the cross section predicted at $32$ GeV,
$\approx 6$ nb is not so much less then
that observed in [4] for C-production by pions, $40$ nb whereas, naively,
one would expect a suppression by more than three orders of magnitude
from the factor $4\pi\alpha/f_{\phi}^2$ in (2). Such a
small
cross section of hadroproduction means strong suppression, by
more than two order of magnitude, of the $C\rightarrow\pi\pi$
decay, if $C$ is really a diquark-antidiquark state. Such a
suppression can better be understood in the time-reversed
channel, $q\bar q\rightarrow C$. Indeed, one of the standard decays of a
$q\bar q$ state would, for instance, be the creation of an additional
$s\bar s$ pair from the vacuum; this would result in the production of a
$K\bar K$ pair. By contrast, a $\phi\pi$ final state is excluded for such
a mechanism due to OZI suppression. The mechanism is the same as for baryon
pair production in the color flux model suggested in [8]. If the
pair
$q\bar q$ created from the vacuum has the ``wrong" color,
{\it i.e.} different from the parent $q\bar q$ color, it does not
completely screen the external color field, and a diquark-
antidiquark state is produced.
If this state is below threshold for $B \bar B$ production, it can only
decay by means of string rearrangement which is what the C-meson is
supposed to do.

The production of a
``wrong" color quarks is estimated [8] to be suppressed by an
order of
magnitude in comparison with a ``right" color $q\bar q$ pair. This,
however, does not seem enough to explain numerically the dramatic
suppression of the $2\pi$ decay mode of the C-meson.
\smallskip

Another mechanism of C-meson photoproduction which we
should discuss briefly
is the one originally suggested in [3], {\it i.e.} the direct
transition $\gamma\rightarrow C$ (fig.4). The vertex of such a
transition is unknown, and in [3] it was assumed to be smaller than
$\gamma\rightarrow\rho$ by an order of magnitude. However,
a photon can convert directly only to a $q\bar q$ pair, and the
subsequent transition from a $q\bar q$ system to the C-meson proves to be
suppressed by more that two orders of magnitude, as
mentioned
above. Thus we estimate the cross section of photoproduction of the C-meson
by the mechanism of fig.4 at about $2-3$ nb. This contribution is
negligibly small in comparison with one under consideration in the
few GeV energy range. Being independent of energy, however, it can
become
important
at high energies.
\smallskip

It worth mentioning also a last source of background to $\phi\pi$
photoproduction which comes from the quasielastic production of
$\phi$ accompanied by diffractive dissociation of the target
proton to $p\pi$.
This can be easily estimated. The cross section
for elastic photoproduction of $\phi$ in a few GeV energy
range is about $0.4-0.5\;\mu$b
[6]. Diffractive dissociation
cross section, for
instance $pp\rightarrow pX$ is about an order of magnitude less
than the elastic one in this energy range. Only 1/3 of it goes
to $p\pi^0$. Assuming
factorization, we can estimate the contribution of the term under discussion
to about $13$ nb. This is quite small as compared with the cross
section shown in fig.2. Besides, this background can be substantially
reduced with proper geometry of the experiment. Although the cross section of
this process is energy independent, its contribution to the mass
interval of $\phi\pi$ under discussion steeply vanishes with
energy.
\smallskip

{\it 5. Production of C-meson by kaons}: Since the
diquark-antidiquark structure of C-meson supposes a branching ratio
$C\rightarrow K\bar K$ about the same as $C\rightarrow\phi\pi$,
one can consider another process of unsuppressed production of
$C$: $KN\rightarrow\phi\pi\Lambda$. The corresponding diagram is
shown in fig.5. Assuming that $\Gamma_{\bar KK}\approx
1/2\Gamma_{tot}$, and $g_{\pi N\Lambda}\approx g_{\pi NN}$
(according to $SU_3$), we get the cross section of C production
shown in fig.6. It proves to be quite large, about $0.1$ mb at
the maximum which is around $4-5$ GeV kaon energy.

{\it 6. Conclusions}: The above estimates for the $C(1480)$
production cross sections by photons and kaons, contain no free
parameters. The only assumption made is the diquark-antidiquark
structure of the C-meson. Therefore, if an experiment will fail to
observe the signal at the predicted level, it will reject the
above interpretation of the C-meson, or its existence at all. On the
other hand, in the case of a successful measurement one gets a
direct information about branchings of C-decay to $\phi\pi$ and
$K\bar K$.

In closing, it is interesting that the issue raised in this paper
should readily be solved in the forthcoming search at CEBAF
in the proposed measurement of rare radiative decays of the $\phi$
meson, where, with a 4 GeV photon beam of intensity $5 \times 10^7$ sec
a branching ratio sensitivity for $\phi$ decays of about
$10^{-5}$ is expected [9].
\bigskip

\noindent{\bf Acknowledgements.} The authors
greatly appreciate very useful and stimulating
discussions with Alex Dzierba and with Don Lichtenberg.
B.Z.K.	thanks Steven Vigdor and the members of the  IUCF of Indiana
University for their hospitality while this work was done.

\vfill\eject
 \centerline{\bf References.}

\begin{enumerate}

\item  D. B. Lichtenberg and L. J. Tassie; Phys. Rev. {\bf 155}
(1967), 1601.

\item  M. Anselmino, S. Ekelin, S. Fredriksson, D. B. Lichtenberg and
E. Predazzi; Rev. Mod. Phys. {\bf 65} (1993), 1199.

\item  F. E. Close and H. J. Lipkin; Phys. Rev. Lett. {\bf 41} (1978)
1263; see also, by the same authors, Phys. Lett. {\bf B 196} (1987) 245.

\item  S. I. Bityukov {\it et al.}; Phys. Lett. {\bf B 188} (1988) 383.

\item  J Reifenr\"other {\it et al.}; Phys. Lett. {\bf B 267} (1991) 299.

\item  T. H. Bauer, R. T. Spital and D. R. Yennie; Rev. Mod. Phys. {\bf 50}
(1978) 261.

\item M. Bishari; Phys. Lett. {\bf B 38} (1972) 510.

\item  A. Casher, H. Neuberger, and S. Nussinov, Phys. Rev. {\bf
D 20} (1979) 179

\item  H. Crannell {\it et al.}; Letter of Intent to the Continuous Beam
Accelerator Facility for {\it Measurement of Rare Decays of the $\phi$
Meson}, Indiana Univ. Dec. 1993.
\end{enumerate}
\bigskip
\bigskip

\centerline{\bf Figure Captions.}

Fig. 1. Relevant diagram for C-meson photoproduction.

Fig. 2. Energy dependence of the C-meson photoproduction cross
section.

Fig. 3. Mass distribution for C-meson photoproduction at
$E_\gamma = 4$ GeV.

Fig. 4. C-meson photoproduction via direct $\gamma \rightarrow C$
transition.

Fig. 5. C-meson production by kaons.

Fig. 6. Energy dependence of the C-meson cross section for production
by kaons.

\end{document}